\documentclass[%
superscriptaddress,
preprint,
preprintnumbers,
 amsmath,amssymb,
 aps,
floatfix,
]{revtex4-2}

\usepackage{color,soul}
\usepackage{graphicx}
\usepackage{dcolumn}
\usepackage{bm}
\usepackage{float}

\newcommand{\VO}{VO\textsubscript{2} }
\newcommand{\VOnospace}{VO\textsubscript{2}}
\newcommand{\textapprox}{$\sim$}

\begin{document}


\title{High-Resolution Full-field Structural Microscopy of the Voltage Induced Filament Formation in Neuromorphic Devices}

\author{Elliot Kisiel}
\affiliation{Physics Department, University of California - San Diego, 9500 Gilman Dr, La Jolla, CA 92093, USA.}
\affiliation{X-ray Science Division, Argonne National Laboratory, 9700 S Cass Ave, Lemont, IL 60439, USA}
\author{Pavel Salev}
\affiliation{Physics Department, University of California - San Diego, 9500 Gilman Dr, La Jolla, CA 92093, USA.}
\affiliation{Department Physics and Astronomy,  University of Denver, 2199 S University Blvd, Denver, CO 80210, USA.}
\author{Ishwor Poudyal}
\affiliation{X-ray Science Division, Argonne National Laboratory, 9700 S Cass Ave, Lemont, IL 60439, USA}
\affiliation{Materials Science Division, Argonne National Laboratory, 9700 S Cass Ave, Lemont, IL 60439, USA}
\author{Fellipe Baptista}
\affiliation{Los Alamos National Laboratory, Los Alamos, New Mexico 87544, USA.}
\affiliation{Centro Brasileiro de Pesquisas Físicas, 22290, Rio de Janeiro, RJ, Brazil.}
\author{Fanny Rodolakis}
\affiliation{X-ray Science Division, Argonne National Laboratory, 9700 S Cass Ave, Lemont, IL 60439, USA}
\author{Zhan Zhang}
\affiliation{X-ray Science Division, Argonne National Laboratory, 9700 S Cass Ave, Lemont, IL 60439, USA}
\author{Oleg Shpyrko}
\affiliation{Physics Department, University of California - San Diego, 9500 Gilman Dr, La Jolla, CA 92093, USA.}
\author{Ivan K. Schuller}
\affiliation{Physics Department, University of California - San Diego, 9500 Gilman Dr, La Jolla, CA 92093, USA.}
\author{Zahir Islam}
\affiliation{X-ray Science Division, Argonne National Laboratory, 9700 S Cass Ave, Lemont, IL 60439, USA}
\author{Alex Frano}
\affiliation{Physics Department, University of California - San Diego, 9500 Gilman Dr, La Jolla, CA 92093, USA.}

\date{\today}

\begin{abstract}

Neuromorphic functionalities in memristive devices are commonly associated with the ability to electrically create local conductive pathways by resistive switching. 
The archetypal correlated material, \VOnospace, has been intensively studied for its complex electronic and structural phase transition as well as its filament formation under applied voltages. 
Local structural studies of the filament behavior are often limited due to time-consuming rastering which makes impractical many experiments aimed at investigating large spatial areas or temporal dynamics associated with the electrical triggering of the phase transition. 
Utilizing Dark Field X-ray Microscopy (DFXM), a novel full-field x-ray imaging technique, we study this complex filament formation process \textit{in-operando} in \VO devices from a structural perspective. 
We show that prior to filament formation, there is a significant gain of the metallic rutile phase beneath the metal electrodes that define the device. 
We observed that the filament formation follows a preferential path determined by the nucleation sites within the device. These nucleation sites are predisposed to the phase transition and can persistently maintain the high-temperature rutile phase even after returning to room temperature, which can enable a novel training/learning mechanism. 
Filament formation also appears to follow a preferential path determined by a nucleation site within the device which is predisposed to the rutile transition even after returning to room temperature. 
Finally, we found that small isolated low-temperature phase clusters can be present inside the high-temperature filaments indicating that the filament structure is not uniform. 
Our results provide a unique perspective on the electrically induced filament formation in metal-insulator transition materials, which further the basic understanding of this resistive switching and can provide guidelines for engineering switching devices with specific properties.

\end{abstract}

\maketitle


\section{Introduction}

In recent years, the development of neuromorphic devices has gained significant attention as a promising approach toward creating advanced energy-efficient computational systems that mimic the complex functionalities of the human brain \cite{NES, shriram, bioPlaus, criticality, challenges, memristReview, RN102, RN103, RN106, RN69, RN71, RN83}. 
Vanadium dioxide, \VOnospace, is a promising material platform to implement neuromorphic devices because of its intriguing electronic and structural properties, particularly the ability of \VO to undergo a metal-to-insulator transition (MIT) coupled with a monoclinic to rutile 1st order structural phase transition \cite{VO2original, isostructual, monoclinicDomains, vo2review}. 
\VO has a moderate transition temperature of T\textsubscript{c} \textapprox 340 K, i.e., just above room temperature, and this phase transition can be induced by various stimuli including optical, electrical, and mechanical perturbations \cite{metalfilamentdomains, opticalChangesFast, opticalPropertiesTransition,  spatiotemporal, ThzIMT, memoryDynamics, memoryResistive, opticalDisorder, CdSVO2}. 
These unique characteristics have paved the way for the implementation of \VO memristive and signal processing devices into neuromorphic systems \cite{workingNetwork, neuralNetwork}.

Electrical triggering of the MIT in \VO devices, which produces volatile resistive switching, is a highly active research area. 
Multiple studies have shown the resistive switching to be a result of a metallic filament forming between the electrodes of the device \cite{spatiotemporal, vo2nanodiffraction, metalfilamentdomains, subthreshold, VO2TEM}. 
Investigations of this filamentary behavior usually probe only the electronic transition while structural microscopy studies are often limited in scope due to long raster times (e.g., x-ray micro- and nanodiffraction) or due to destructive specimen preparation procedures (e.g., transmission electron microscopy) \cite{spatiotemporal, imageFil, metalfilamentdomains, subthreshold, VO2TEM, vo2nanodiffraction}. 
The coupling of structural and electronic degrees of freedom in correlated oxides such as \VO can give rise to exotic phases, therefore studying the structural properties of the voltage-induce filament formation is important to gain a basic understanding of the MIT switching process \cite{M2Phase}. 
Utilizing x-ray techniques provides the ability to directly explore the structural transition that accompanies the electrical MIT triggering. 
Furthermore, x-ray studies allow investigation of the switching process in the entire device structure, for example, probing the structural changes in the areas beneath the electrodes, which is often inaccessible using techniques such as optical microscopy.

The coupling between the MIT and structural transition in \VO gives an opportunity to explore \textit{in-operando} the structural signature of electrical MIT triggering \cite{isostructual, vo2Phase}. 
Although the question, if these two transitions always coincide or if they can be decoupled still remains open, it is usually accepted that under normal electric device operation (i.e. without using ultra-fast optical pumps), the MIT and monoclinic-rutile transition go hand-in-hand \cite{VO2MM1, fastdynamicMMPhase}. 
A previous x-ray nanodiffraction study has utilized these coupled transitions to probe the switching in \VO devices, however, nanodiffraction imaging required time-intensive rastering resulting in several hours data acquisition time \cite{vo2nanodiffraction}. 
The time-intensive rastering makes it impractical to explore the voltage cycling effects in the switching devices and makes the studies of the filament relaxation after turning off the driving voltage unreliable as the x-ray signal in different regions is acquired at different times. 
To address such challenges, two key experimental improvements are necessary: (i) an x-ray imaging technique with a large field of view (FOV) to acquire the entire device image in a single shot and (ii) short exposure time operation. 
A characterization technique that implements the above requirements is indispensable to gain a comprehensive understanding and optimize the practical performance of MIT switching devices.

Dark field x-ray microscopy (DFXM) recently emerged as a powerful tool for studying material local structural properties. 
DFXM employs the spatially resolved detection of scattered x-rays from a sample, providing enhanced contrast and sensitivity to subtle mesoscopic structural changes \cite{DFXM109, DFXM4, DFXM58, DFXM59}. 
The first DFXM studies have been applied to bulk crystals and relatively thick films with high-Z elements, which prompts strong x-ray diffraction intensities \cite{BFOFilms, MCTFilm, DFXM109}.  
\VOnospace, being composed of lower Z elements, proves challenging to DFXM due to its low scattering efficiency. 
The low-scattering-intensity problem is further exacerbated in thin films due to the reduced volume. 
In this work, we overcame these challenges. 
We present routes that can help apply DFXM to other systems where the x-ray scattering is weak, for example, scattering from charge density waves.

Although multiple works have explored electronic features of the filament formation in \VOnospace, many open questions remain about why filaments form preferentially in specific locations, about the memory/training effects that emerge during repeated voltage cycling, etc.  
Our in-operando DFXM experiments in \VO switching devices enabled the study of the intricate morphological properties of the material at the nanoscale. 
The full-field imaging capability of DFXM allowed for the simultaneous acquisition of large-area, high-resolution ($<$ 100 nm) images, permitting a comprehensive analysis of the filament formation and relaxation. 
The fast signal acquisition times afforded by DFXM permitted us to perform voltage cycling studies which are key to understanding the MIT switching repeatability and emerging memory/training phenomena. 
Specifically, we observed that filaments in \VO devices have local regions of monoclinic phase intermixed with the rutile filament. 
We found that the parts of \VO switch into the rutile phase underneath the electrodes prior to the filament formation providing a possible explanation for why the filament appears in a specific place. 
After repeatedly cycling the device through the MIT switching, we observed that certain regions become predisposed to switching into the rutile phase and that this predisposition persists after cooling \VO to room temperature. 
Such regions act as nucleation points for filament formation and can possibly be structurally engineered to improve the switching repeatability and energy efficiency in MIT memristive devices.

\begin{figure*}
\includegraphics[width=\textwidth]{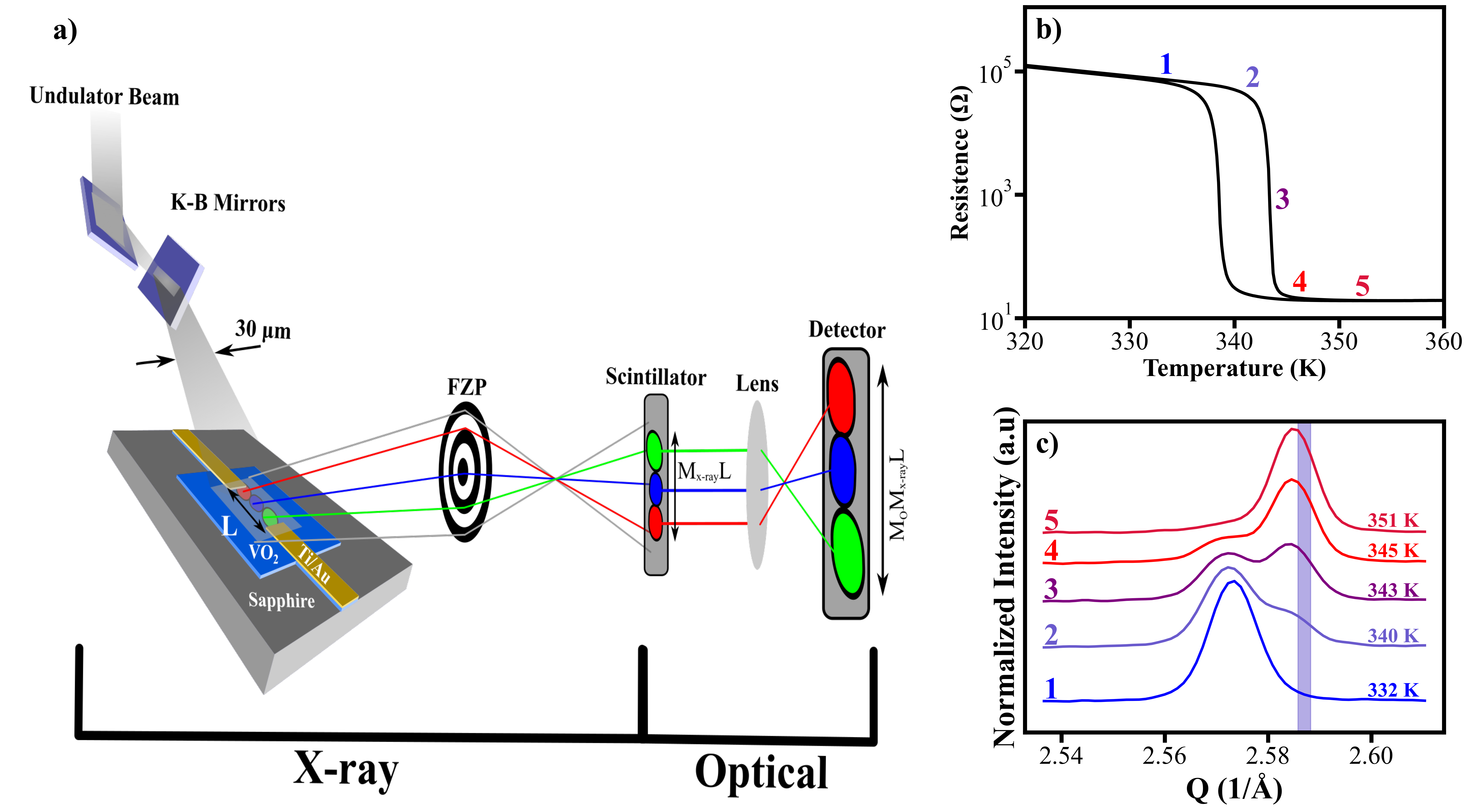}
\caption{\label{fig:dfxmSetup} a) The setup of the DFXM showing the use of a Fresnel Zone Plate (FZP). 
The focused beam from the KB mirrors before the sample increases the flux density in the region of interest. 
Spatially separated regions (red, blue, and green ovals) that are diffracting at the same angle, when passed through the FZP, are mapped to spatially different regions on the detector. 
We utilized an indirect detection scheme to further improve the magnification by a factor of 20 (M\textsubscript{O}). 
b) The temperature hysteresis associated with the cycled device. 
Label points correspond with the temperatures where (c) was collected showing a coincidence of the electrical and structural measurements. 
c) The temperature evolution of the structural phase transition from monoclinic (blue) to rutile (red). 
The shaded grey region marks the scattering condition where the images in Fig. \ref{fig:cycling} were recorded.}
\end{figure*}
\section{Methods and Materials}

300-nm-thick \VO film was grown by the reactive rf sputtering using a stoichiometric V\textsubscript{2}O\textsubscript{3} target. 
Details about the film growth and device preparation can be found elsewhere \cite{spatiotemporal}. 
The \VO phase formation was confirmed by the specular x-ray diffraction measurements (Suppl. Fig. S1b). 
(20 nm Ti)/(100 nm Au) electrodes defining the planar 40×160 \textmu m\textsuperscript{2} and 10×40 \textmu m\textsuperscript{2} two terminal devices were prepared using the standard optical lithography and e-beam evaporation (Suppl. Fig. S2). 
Resistance-temperature measurements showed an abrupt 1st-order phase transition at T\textsubscript{c} $\sim$ 340 K with roughly 4 orders-of-magnitude resistance change (Fig. \ref{fig:dfxmSetup}b). 
Importantly, the resistance-temperature curves of the film before and after device fabrication were identical (Suppl. Fig. S1a), which indicates that the device preparation procedures did not alter the \VO film's stoichiometry or structure.
Two devices were studied: a 40 x 160 \textmu m\textsuperscript{2}, hereafter device 1, and a 10 x 40 \textmu m\textsuperscript{2}, hereafter device 2. 
A filament on device 1 was achieved by heating the \VO sample past its transition, applying 10 V, and then removing the heat source allowing a filament to form which was subsequently stepped down in voltage (Suppl. Fig. S6a). 
Voltage cycling of device 2 followed the procedure of 1), the sample was maintained at 330K while ramping up the voltage to 10V over \textapprox 15 minutes; 2), after reaching 10 V, the sample was returned to room temperature; 3), the voltage was ramped down to 0 V within \textapprox 5 minutes (Suppl. Fig. S6b). 
Each voltage cycle took approximately 30 minutes to complete and when returning back to 0 V, the sample was allowed to thermalize back to room temperature for 3 minutes. 
This is in contrast to most other voltage cycles performed on \VO devices which typically maintain a constant system heat while cycling the voltage \cite{spatiotemporal, subthreshold}.

Fig. \ref{fig:dfxmSetup}a shows the DFXM setup that we used to explore local structural changes during the electrically induced filament formation in \VO devices.  
The measurements were performed at 6-ID-C and 33-ID-D beamlines of The Advanced Photon Source (APS)  \cite{XRIM1,XRIM2, DFXM109, IDC}. 
We utilized dark-field x-ray microscopy (DFXM) which provides a spatial mapping of the diffracting regions in the sample to study the mesoscale structures of these \VO devices. 
Magnification in the DFXM experiments is achieved by placing an objective lens in the path of the diffracted beam. 
Similar to optical microscopy, the objective lens refracts the diffracted x-ray beam producing a magnified device image on the scintillating detector. 
Unlike optical microscopy which relies on acquiring the reflected light, the contrast in DFXM originates from diffraction, i.e., the images acquired by the detector correspond to spatial variations in the diffraction conditions. 

DFXM operates via an objective lens placed in the path of the diffracted beam (Fig. \ref{fig:dfxmSetup}a) which allows for diffraction contrast originating from the distinct parts of the sample.  
We utilized a Fresnel Zone Plate (FZP) with a 0\textsuperscript{th} order block (not shown) as an objective lens with a focal length of 56.5 mm at 10 keV, (100 \textmu m diameter and 70 nm outer zone width). 
With a sample-to-scintillator distance of 1470 mm, we achieved an overall x-ray magnification M\textsubscript{x-ray}\textapprox24. 
The short focal length offered by the FZP allows for this high magnification and high resolution necessary to image nanoscale domains of the monoclinic (M1) and rutile (R) phases. 
An additional optical magnification M\textsubscript{O}=20 coming from an optical lens placed between the x-ray scintillator and optical detector further enhanced the magnification to \textapprox 480. 
Overall, we achieved an effective pixel size of 15 nm and an approximate resolution of 70 nm, on the same order of magnitude as the nanodiffraction resolution of 25 nm\cite{vo2nanodiffraction}. 
The main advantage of the DFXM, however, is the ability to record full-field-view images in a single shot without the need to raster the x-ray beam, unlike in the nanodiffraction approach.

One of the challenges of performing DFXM on \VO is the low intensity of the diffracted beam because of the low-Z element composition, vanadium and oxygen. 
Preliminary experiments at 6-ID-C in which the sample was illuminated by an unfocused 20 keV x-ray beam yielded discernable, but weak-contrast images (Suppl. Fig. S4) \cite{directDetection}. 
At 33-ID-D beamline, we improved the signal from the film by utilizing a pair of Kirkpatrick-Baez (KB) mirrors to focus the beam to a 30 x 30 \textmu m\textsuperscript{2} spot size to increase the incident flux density (\textapprox 10\textsuperscript{13} photons/s) in the region of interest while maintaining a full device view (Fig. \ref{fig:dfxmSetup}a).  
We note that the purpose of KB mirrors in our DFXM setup differed from the conventional microdiffraction. 
In microdiffraction, the beam is focused by KB mirrors to a small spot size and the imaging is achieved by rastering the beam over the sample area and acquiring local diffraction patterns. 
In our DFXM experiments, the KB mirrors were focusing the beam to increase the incident x-ray flux density that illuminated the entire device area at once. 
We further improved the resolution along the x-ray propagation direction by reducing the beam energy to 10 keV.
We measured the \VO film in symmetrical reflection geometry at an incident angle of \textapprox 14.8\textdegree\ resulting in a beam footprint of roughly 30 x 120 \textmu m\textsuperscript{2}, which matches our field of view well. 
Accordingly, the resolution along the x-ray direction changes by a factor of 4 to about 300 nm.

The contrast mechanism of DFXM originates from diffraction which allows for the spatial differentiation between the R and M1 phases in the \VO by collecting images at their respective Bragg conditions. 
Fig. \ref{fig:dfxmSetup}c shows the equilibrium (V = 0) structural evolution of \VO across the phase transition. 
The separation of the momentum transfer, $Q$, between the R (high temperature) and M1 (low temperature) Bragg peaks is significant enough to image the two distinct phases and their spatial distribution that emerges when voltage is applied to \VO devices to trigger the MIT. 
When studying the phase distribution and evolution (Fig. \ref{fig:COM}b), a series of images were taken at each voltage along a line in reciprocal space that intersects both the M1 and R diffraction peaks.  
For multiple cycle measurements (Fig. \ref{fig:cycling}b \& \ref{fig:cycling}c), images were collected at the right shoulder of the R peak depicted by the grey region in Figure \ref{fig:dfxmSetup}c. 
At this angle corresponding to the Bragg peak shoulder, the R phase is the dominant contributor. 
 
\section{Results and Discussions}

\begin{figure*}
\includegraphics[width=\textwidth]{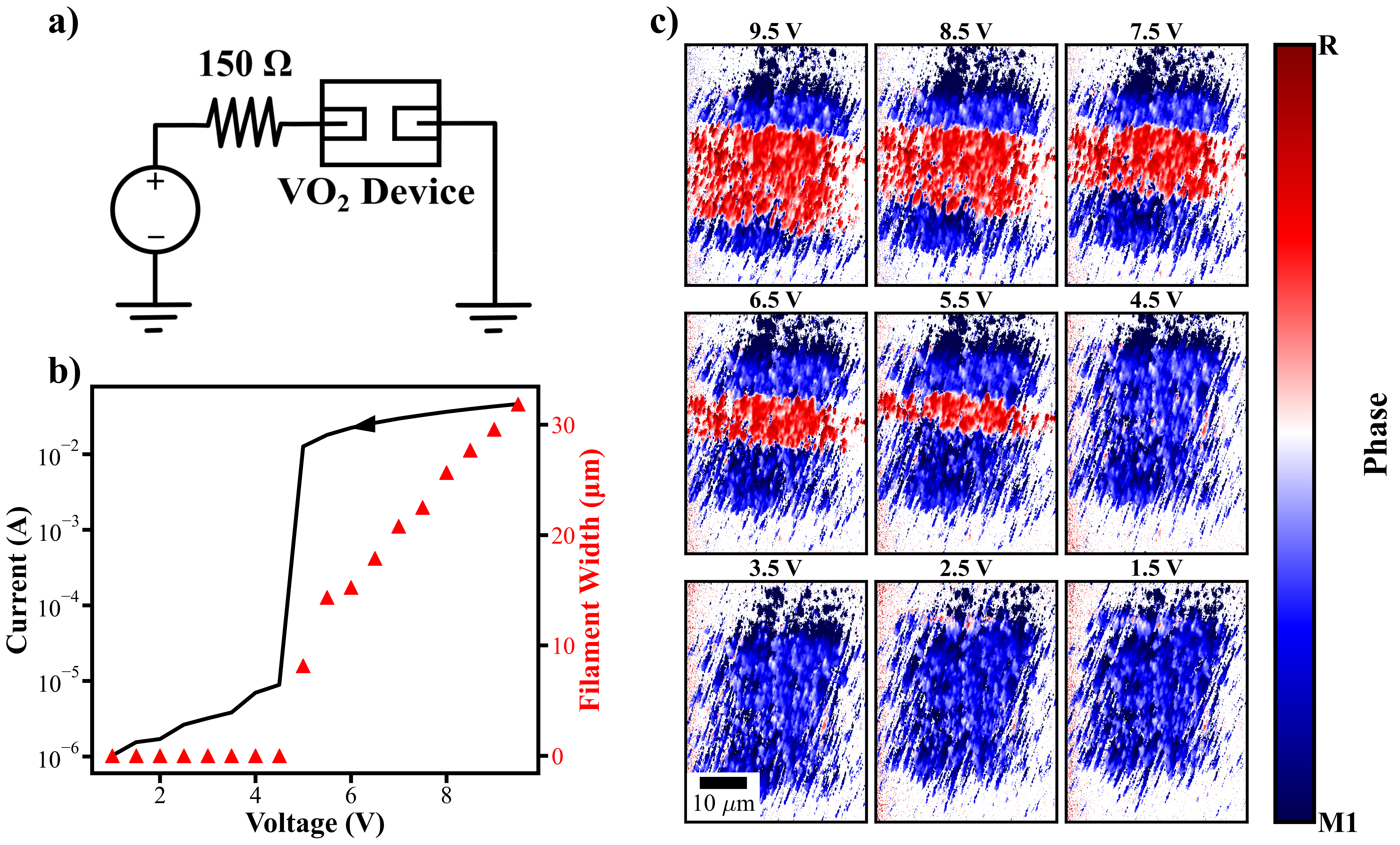}
\caption{\label{fig:COM} a) The voltage-current response of the \VO device as the voltage is reduced on device 2 (black line). 
The filament width as the voltage is reduced which follows a linear decrease as power is reduced (red triangles). 
b) Maps of the filament region in device 2 with their corresponding voltages above each image. 
Each of these maps was reconstructed from a series of images collected along a line in reciprocal space that intersects both the M1 and R diffraction peaks. 
Pixel-wise analysis of the images yields two groups of the calculated center of mass (COM) in terms of $Q$, which correspond to the R phase (the red regions) and the M1 phase (the blue regions) \cite{darfix}. 
The electrodes for the device are just outside the field of view on the left and right sides.
}
\end{figure*}

DFXM imaging showed the R phase filament formation in the M1 phase matrix during the electrical MIT switching in \VO devices (Fig. \ref{fig:COM}). 
A voltage source was used to drive the switching and a 150 Ω resistor was connected in series with the \VO device to limit the current after the MIT triggering (Fig. \ref{fig:COM}a).
The measurements shown in Fig. \ref{fig:COM} were performed on device 1 while the voltage stepped from 10 V to 0 V at room temperature. 
A series of images were taken at each voltage along a line in reciprocal space that intersects both the M1 and R diffraction peaks.  
For each pixel, its intensity varies as reciprocal space position, i.e., $Q$, changes. 
The calculated center of mass (COM) in terms of $Q$ for each pixel was obtained and they congregated into two groups, corresponding to the M1 and R phase peak positions, respectively.  
Color is assigned to each pixel based on its COM position. 
The red regions in the maps indicate the presence of the R phase at that location while the regions that are blue correspond to the M1 phase.
At the switching threshold voltage, V\textsubscript{th} \textapprox 5 V, the current through the \VO device abruptly drops indicating that the device suddenly became insulating (Fig. \ref{fig:COM}b, black curve).  
At V\textsubscript{th} and above, DFXM imaging reveals a clear filamentary phase distribution in the device: an R phase channel is surrounded by the M1 phase matrix. 
After device 1 returns to its insulating phase as revealed by electrical measurements, the observed device area is predominantly in the M1 phase. 
Close to V\textsubscript{th}, however, in the 3.5-5.5 V range, the remnants of a few isolated R phase pockets can be observed within the device area, which is likely due to a local increase in temperature from the near uniform Joule heating of the device \cite{jouleHeating}. 
Assuming the direct correspondence between R/M1 phases and metal/insulator phases, respectively, our DFXM observations of the filament are consistent with the previous optical studies, that is a metallic filament surrounded by the insulating matrix \cite{spatiotemporal, subthreshold, thermal}. 
The filament width in our DFXM experiments is given by the red triangles in Fig. \ref{fig:COM}b. 
The filament becomes wider at higher applied voltages/currents, which is also consistent with previous measurements probing the electrically induced metallic state in \VO \cite{spatiotemporal}. 
The above results provide the first full-field structural observations of the filament formation in \VO and confirm that DFXM is a reliable technique for exploring the structural properties of the voltage-induced phase transition.

While the initial experiments provide confirmation of the DFXM applicability to image the switching in \VOnospace, our measurements further reveal previously unreported features. 
Even though the filament is formed by the R phase, we also observe the presence of isolated M1 phase pockets within the filament region. 
Blue M1 spots can be seen within the red R filament in Fig. \ref{fig:COM}c in 5.5-9.5 V range. 
This coexistence of the M1 phase within the filament indicates that the film has domains that are less prone to switching compared to the remaining parts of the film. 
As the filament is maintained by Joule heating, these M1 domains likely have a higher transition temperature compared with the regions that undergo the MIT switching \cite{thermal}. 
The growth of the filament width with increasing voltage (Fig. \ref{fig:COM}b, red symbols) suggests that the temperature of the filament does not increase as more power is applied to the device, but rather the power goes into switching more regions nearby. 
The underlying structural properties that govern the local \VO domain susceptibility to the electrical switching require further investigation, though this may be related to defects and/or local stoichiometric differences.

\begin{figure}
\includegraphics[width=\textwidth]{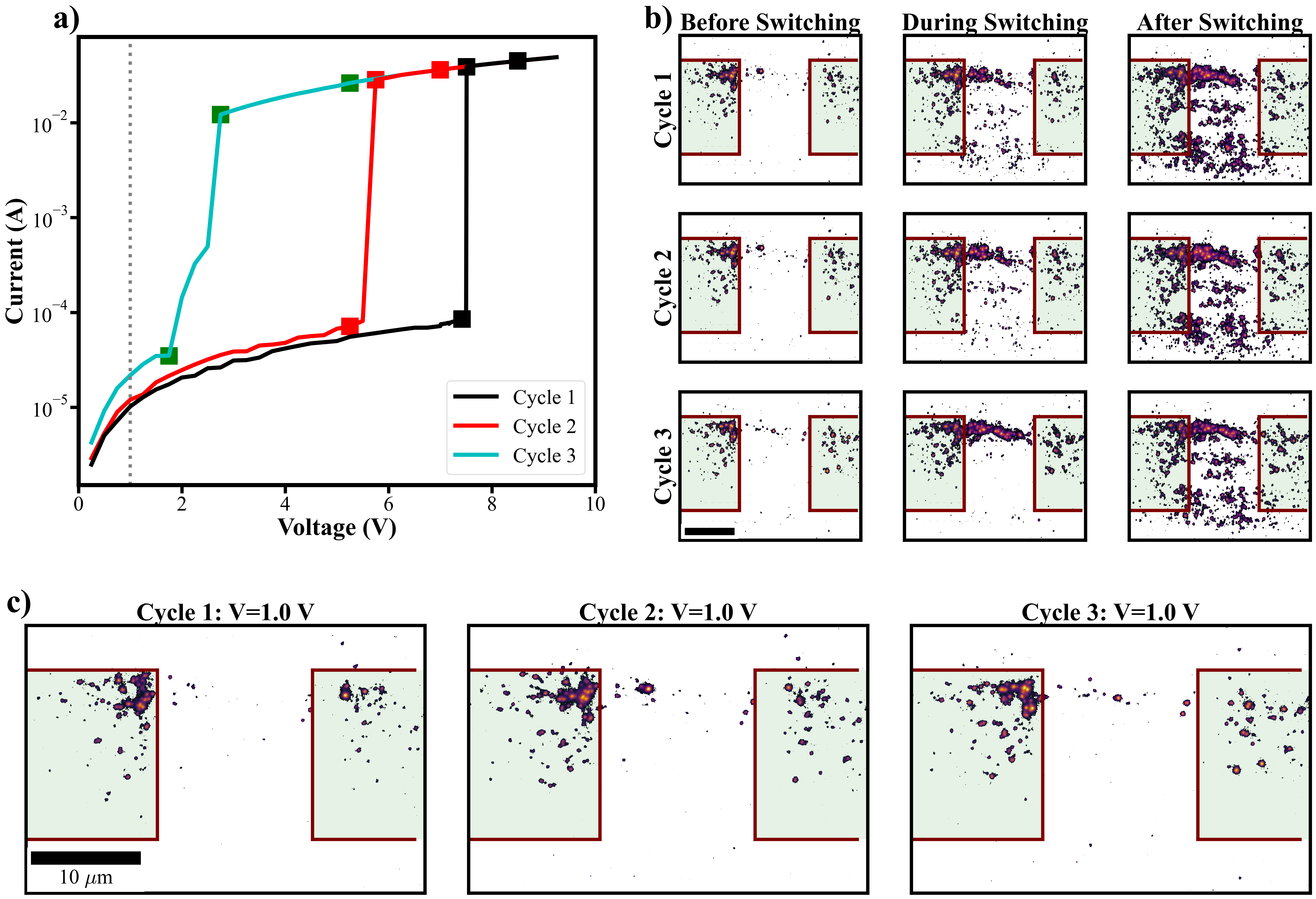}
\caption{\label{fig:cycling} a) Current-voltage curves for the different labeled cycles. 
The square points indicate where the DFXM images were collected. 
The left-, middle, and rightmost points for each respective curve correspond to the before, during, and after switching, respectively in panel (b). 
The grey line is at 1 V indicating the position for (c).  
b) DFXM images collected just before the device switches (first column), during the device switching (second column), and after the device has fully switched (third column) of device 2. 
The same feature shows up during all three cycles on the top of the electrodes indicating a preferential pathway. 
The third cycle shows that this is the primary filament formed during the transition and appears before the other features near the bottom of the electrode. 
The scale bar in the bottom left image corresponds to 10 \textmu m on the sample for all images and electrodes are shown in green. 
c) A comparison of the three cycles at 1 V during the voltage cycles. 
The appearance of the R phase within the device at lower voltages, even after resetting the system back to room temperature, shows there is a local modulation to these nucleation regions which is maintained after this reset.}
\end{figure}

Because of the volatile nature of MIT switching, \VO devices are often used to build spiking oscillators for applications as biologically plausible artificial neurons in neuromorphic hardware  \cite{network1, network2, network3, stoch1, stoch2}.  
Enabling a short/medium-term memory in \VOnospace, i.e., implementing a slowly relaxing switching, might further help to achieve practically useful functionalities such as integrate-and-fire or neuronal adaptation and accommodation operations. 
Our electrical measurements showed that \VO devices can maintain information on medium time scales upon repeated switching cycling (Fig. \ref{fig:cycling}a). 
In this experiment, the switching cycles were performed at 330 K. 
Even though cooling to room temperature can be expected to reset the device phase state, i.e., to convert the \VO volume into M1 phase, we observed a persistent decrease of the threshold voltage (V\textsubscript{th}) upon repeating the switching cycles. 
During the third cycle, we observed \textapprox 80\% reduction of V\textsubscript{th}, from V\textsubscript{th1} \textapprox 7.5 V to V\textsubscript{th3} \textapprox 1.5 V.
We conclude, therefore, that \VO devices can preserve memory about the previous switching events for the duration of at least several minutes.

While electrical measurements directly show a memory effect in \VO switching devices, they do not provide information on the cause of this phenomenon. 
We explored the memory mechanism associated with repeated switching cycling using DFXM (Fig. \ref{fig:cycling}b and \ref{fig:cycling}c). 
Images are collected at the right shoulder of the R peak depicted by the grey shaded region in Fig. \ref{fig:dfxmSetup}c, where the signal is dominantly from R phases, while the M1 phase contribution is negligible. 
As the device was briefly cooled down to room temperature between the electrical cycles, the R phase did not persist through this thermal treatment, i.e., \VO was reset into the M1 phase at the beginning of each switching cycle. 
This indicates that the observed cycle-to-cycle reduction of V\textsubscript{th} is not directly related to the R phase persistence, which potentially could occur as we performed the switching measurements at 330 K, i.e., close to T\textsubscript{C} \textapprox 340 K. 
We observed, however, that during each cycle, the voltage induced R phase filament forms at approximately the same location, close to the top of the device in Fig. \ref{fig:cycling}b. 
While the general filament appearance is nearly identical between the cycles, it can be observed upon closer examination that the R phase domains tend to cluster more closely near the filament as the device undergoes electrical cycling. 
This can be seen as a smaller number of isolated R phase pockets nucleating outside the filament channel with each switching cycle. 
We found that this R phase clustering and the observed cycle-to-cycle decrease of V\textsubscript{th} can be attributed to increasing preferential switching of local regions into the R phase.
Fig. \ref{fig:cycling}c shows the DFXM maps acquired under a small bias voltage of 1V. 
Even though this bias voltage is far below the V\textsubscript{th} of inducing the MIT switching, we observed the nucleation of isolated R phase domains at the locations where the filament eventually forms above V\textsubscript{th}. 
These isolated R phase domains likely serve as current focusing spots that reduce the switching threshold voltage and force the filament formation at the particular location. 
As a control experiment, we performed the DFXM imaging under the same bias voltage but after allowing the device to thermalize for 30 minutes. 
We found that after this 30 min relaxation, the sites that exhibited the low voltage switching have disappeared (Suppl. Fig. S5).
Overall, our experiments directly reveal a regime in which the \VO devices can retain information about the previous switching events on the order of minutes and reset on the order of tens of minutes. 
The intermediate timescales at which DFXM images are collected enables a new perspective on the MIT switching process, thus providing a distinct advantage over other diffraction-based techniques that operate either on hours (micro- and nanodiffraction) or sub-microsecond (ultrafast time-resolved diffraction) timescales.

Previous nanodiffraction measurements showed that the filament in \VO switching devices can partially extend under the electrode \cite{vo2nanodiffraction}. 
Likely because of the small FOV in the nanodiffraction, it was concluded that the proximity to the hot filament is responsible for causing the R phase formation under the electrodes. 
Owing to the wide FOV in our DFXM measurements, we were able to observe that the R phase extends to a very large distance under the electrode, over a 1.5 device gap, where the heating effect of the filament should be insignificant (Fig. \ref{fig:cycling}b). 
A larger area survey of another larger-size device (device 1) showed this filament extension under the electrodes is not just a feature of small-size devices (Suppl. Fig. S7). 
We also observe that this R phase under the electrodes appears prior to filament formation (Fig. \ref{fig:cycling}b, “Before switching” panels). 
This local R phase accumulates within the proximity of the eventual filament formation spot indicating that this R phase under the electrodes may serve as a precursor event to the MIT switching within the device (Fig. \ref{fig:cycling}b). 
The formation of the R phase prior to the filament indicates that the film beneath the electrodes is predisposed to transitioning at lower temperatures compared to the rest of the device. 
Lower transition temperatures of \VO under electrodes can be a result of the electrode fabrication procedures. 
For example, slight chemical composition changes can be due to the exposure of \VO surface to the photoresist developer or due to the reaction between the \VO and Ti adhesion layer in the Ti/Au electrodes. 

\section{Conclusion}

The large FOV and short signal acquisition time of DFXM enabled the observation of unique features of the MIT switching in \VO devices.  
We found that (i) R phase forms in the large extended areas beneath the electrodes prior to device switching, (ii) isolated R phase domains can nucleate within the device resulting in a preferential filament pathway, and (iii) MIT switching induces a medium-term memory upon repeated electrical cycling resulting in the nucleation of R phase domains at applied voltages much lower than the switching threshold.  
This memory effect persists even after a short thermal reset of the device (a longer thermal reset, however, erases the memory) and prompts filament formation along a preferential pathway. 
The identification of a preferential pathway of filament formation suggests that local structural modulation could be used to entice filament formation in a desired place and can help reduce the activation voltages. 
Such modulation could be engineered using laser-etching, local hydrogen doping, or focused ion beam irradiation \cite{VO2hydrogen, FIB}. 
The formation of the rutile phase beneath the electrodes appearing prior to device switching and extending well beyond the local influence of the filament indicates a complex interaction between the electrodes and film.  
Further studies of the electrode-film interactions can have a significant impact on the device construction and network integration to achieve controlled local heating of the electrodes resulting in switching voltage reduction and improved energy efficiency. 
With the continuing development of the synchrotron radiation sources and detector capabilities, observations of the structural dynamics at timescales of milliseconds and below using DFXM imaging may be possible \cite{directDetection, APS-U_upgrade}. 
Our results obtained in \VO thin film devices also provide a benchmark for the DFXM studies of other low-intensity scattering phenomena, such as scattering from magnetic features and charge density waves.

\begin{acknowledgments}

This work was supported as part of the "Quantum Materials for Energy Efficient Neuromorphic Computing" (Q-MEEN-C), an Energy Frontier Research Center funded by the U.S. Department of Energy, Office of Science, Basic Energy Sciences under the Award No. DESC0019273.
This research used resources of the Advanced Photon Source, a U.S. Department of Energy (DOE) Office of Science user facility operated for the DOE Office of Science by Argonne National Laboratory under Contract No. DE-AC02-06CH11357.
Work at Los Alamos National Laboratory was supported by the U.S. Department of Energy, and the Office of Basic Energy Sciences, Division of Materials Science. 
The authors would like to thank the Karlsruhe Nano Micro Facility (KNMF) for the fabrication of the polymer x-ray optics.

\end{acknowledgments}

\bibliography{VO2_Main}

\end{document}



\title{\textit{Supplementary Materials}: High-Resolution Structural Microscopy of Voltage Induced Filament Formation in Neuromorphic Devices}

\author{Elliot Kisiel}
\affiliation{Physics Department, University of California - San Diego, 9500 Gilman Dr, La Jolla, CA 92093, USA.}
\affiliation{X-ray Science Division, Argonne National Laboratory, 9700 S Cass Ave, Lemont, IL 60439, USA}
\author{Pavel Salev}
\affiliation{Physics Department, University of California - San Diego, 9500 Gilman Dr, La Jolla, CA 92093, USA.}
\affiliation{Department Physics and Astronomy,  University of Denver, 2199 S University Blvd, Denver, CO 80210, USA.}
\author{Ishwor Poudyal}
\affiliation{X-ray Science Division, Argonne National Laboratory, 9700 S Cass Ave, Lemont, IL 60439, USA}
\affiliation{Materials Science Division, Argonne National Laboratory, 9700 S Cass Ave, Lemont, IL 60439, USA}
\author{Fellipe Baptista}
\affiliation{Los Alamos National Laboratory, Los Alamos, New Mexico 87544, USA.}
\affiliation{Centro Brasileiro de Pesquisas Físicas, 22290, Rio de Janeiro, RJ, Brazil.}
\author{Fanny Rodolakis}
\affiliation{X-ray Science Division, Argonne National Laboratory, 9700 S Cass Ave, Lemont, IL 60439, USA}
\author{Zhan Zhang}
\affiliation{X-ray Science Division, Argonne National Laboratory, 9700 S Cass Ave, Lemont, IL 60439, USA}
\author{Oleg Shpyrko}
\affiliation{Physics Department, University of California - San Diego, 9500 Gilman Dr, La Jolla, CA 92093, USA.}
\author{Ivan K. Schuller}
\affiliation{Physics Department, University of California - San Diego, 9500 Gilman Dr, La Jolla, CA 92093, USA.}
\author{Zahir Islam}
\affiliation{X-ray Science Division, Argonne National Laboratory, 9700 S Cass Ave, Lemont, IL 60439, USA}
\author{Alex Frano}
\affiliation{Physics Department, University of California - San Diego, 9500 Gilman Dr, La Jolla, CA 92093, USA.}

\date{\today}

\maketitle


\section{Film Growth and Characterization}

300-nm-thick \VO film was grown by the reactive rf sputtering using a stoichiometric V\textsubscript{2}O\textsubscript{3} target. 
The substrate temperature during the growth was 460$^{\circ}$C. 
The growth was done in 3.4 mTorr 92\% Ar and 8\% O\textsubscript{2} atmosphere. The rf power was 100 W resulting in a $\sim$3.2 nm/min growth rate. After the growth, the sample was cooled down at 12$^{\circ}$C/min to room temperature while maintaining the Ar/O\textsubscript{2} atmosphere. 
To isolate individual devices on the same chip, optical lithography and reactive ion etching were used to pattern the 400×400 \textmu m$^2$ \VO islands surrounding each device. 
The reactive ion etching was done in 50 mTorr 17\% Ar and 83\% Cl\textsubscript{2} atmosphere at 200 W rf power giving a $\sim$100 nm/min etching rate. 

\begin{figure*}
\includegraphics[width=\textwidth]{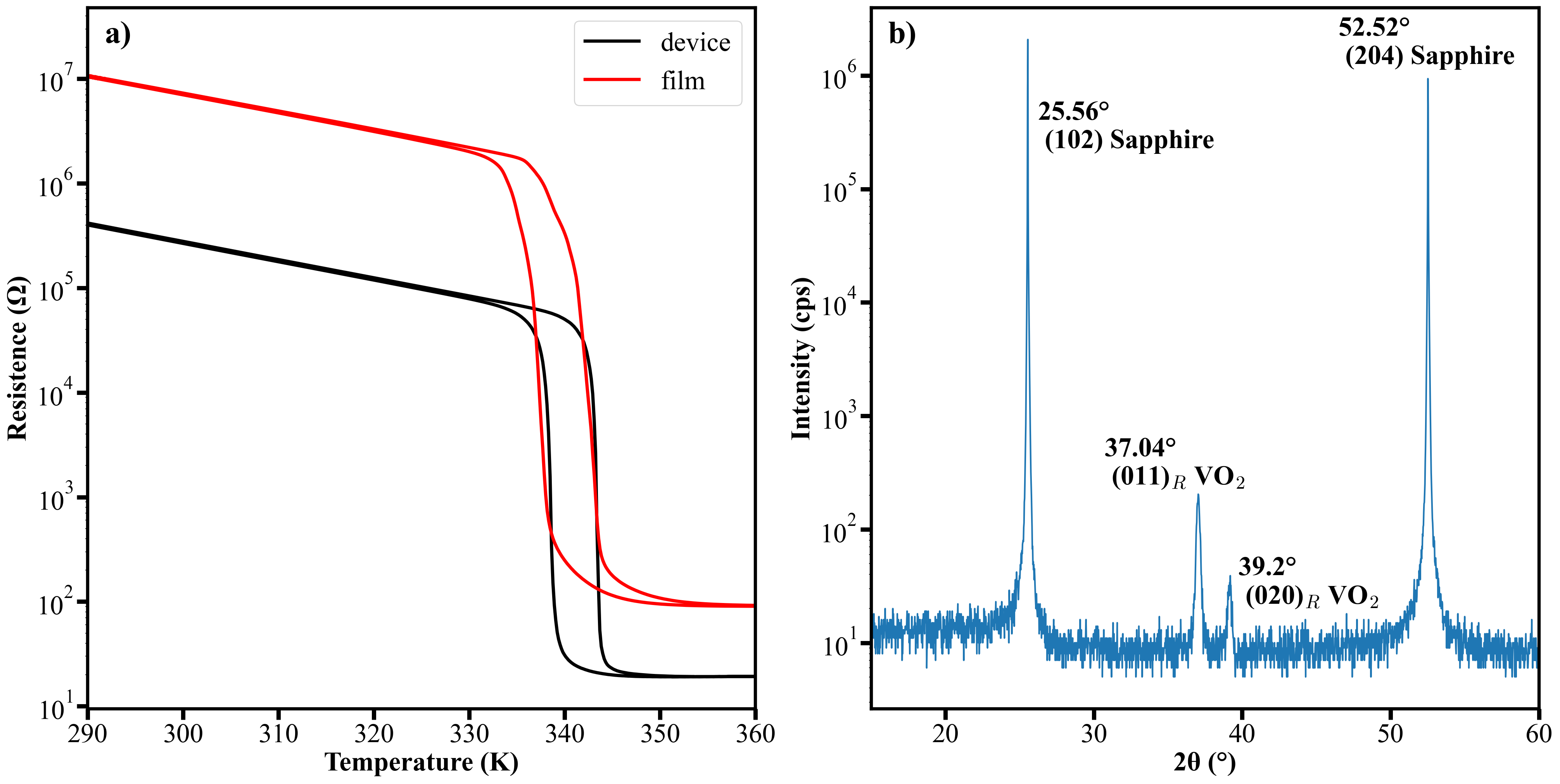}
\caption{\label{supfig:xrd} a) The resistance as a function of temperature showing the hysteresis of the \VO film (red) and etched device (black). 
b) Preliminary specular XRD measurements were taken on the unpatterned film performed at room temperature using a Cu-K$\alpha$ source.}
\end{figure*}

\begin{figure*}
\includegraphics[width=\textwidth]{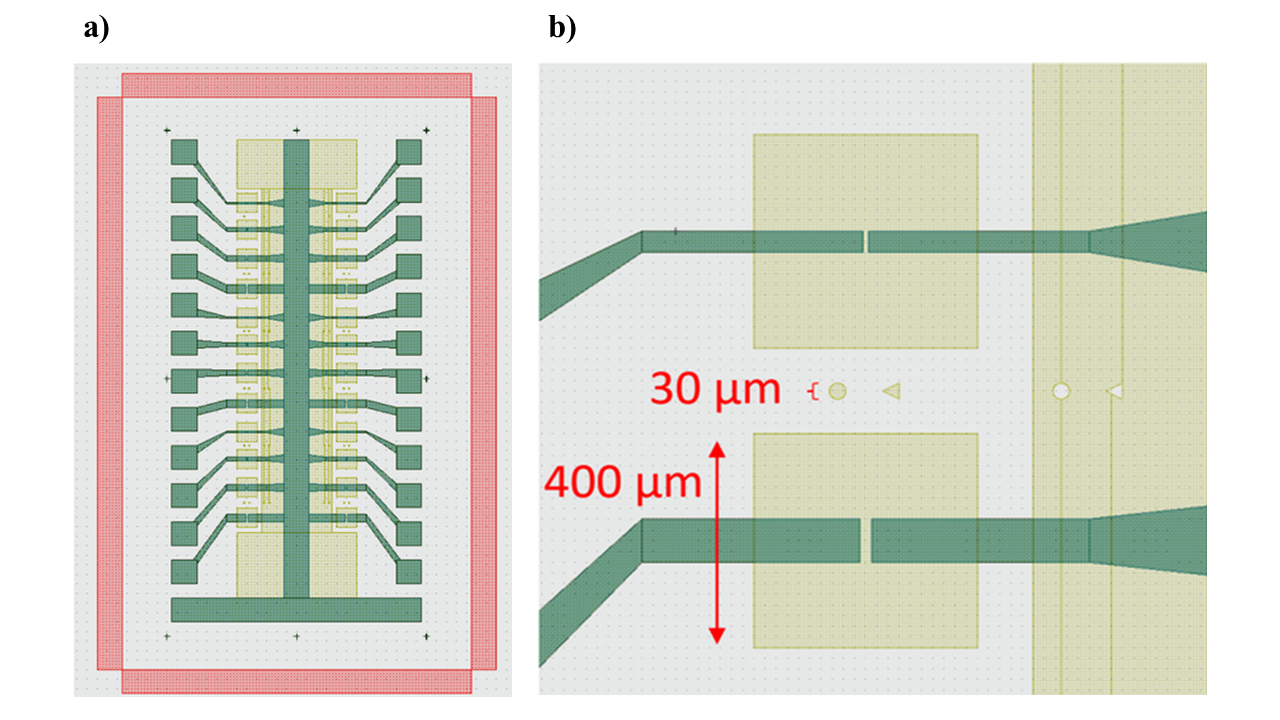}
\caption{\label{supfig:device} a) The overall construction of the devices with pads of \VO film in yellow and Ti/Au electrodes in green. 
A total of 24 devices were made with sizes ranging from 10x40 \textmu m\textsuperscript{2} to 40x160 \textmu m\textsuperscript{2} sized devices. 
b) A zoomed-in view of two of the devices (10x40 \textmu m\textsuperscript{2} and 20x80 \textmu m\textsuperscript{2}) with the 400x400 \textmu m\textsuperscript{2} pad and small fiducials for resolution tests labeled. 
Green indicates the electrodes, yellow the \VO film, and gray indicates bare sapphire.}
\end{figure*}

\begin{figure*}
\includegraphics[width=\textwidth]{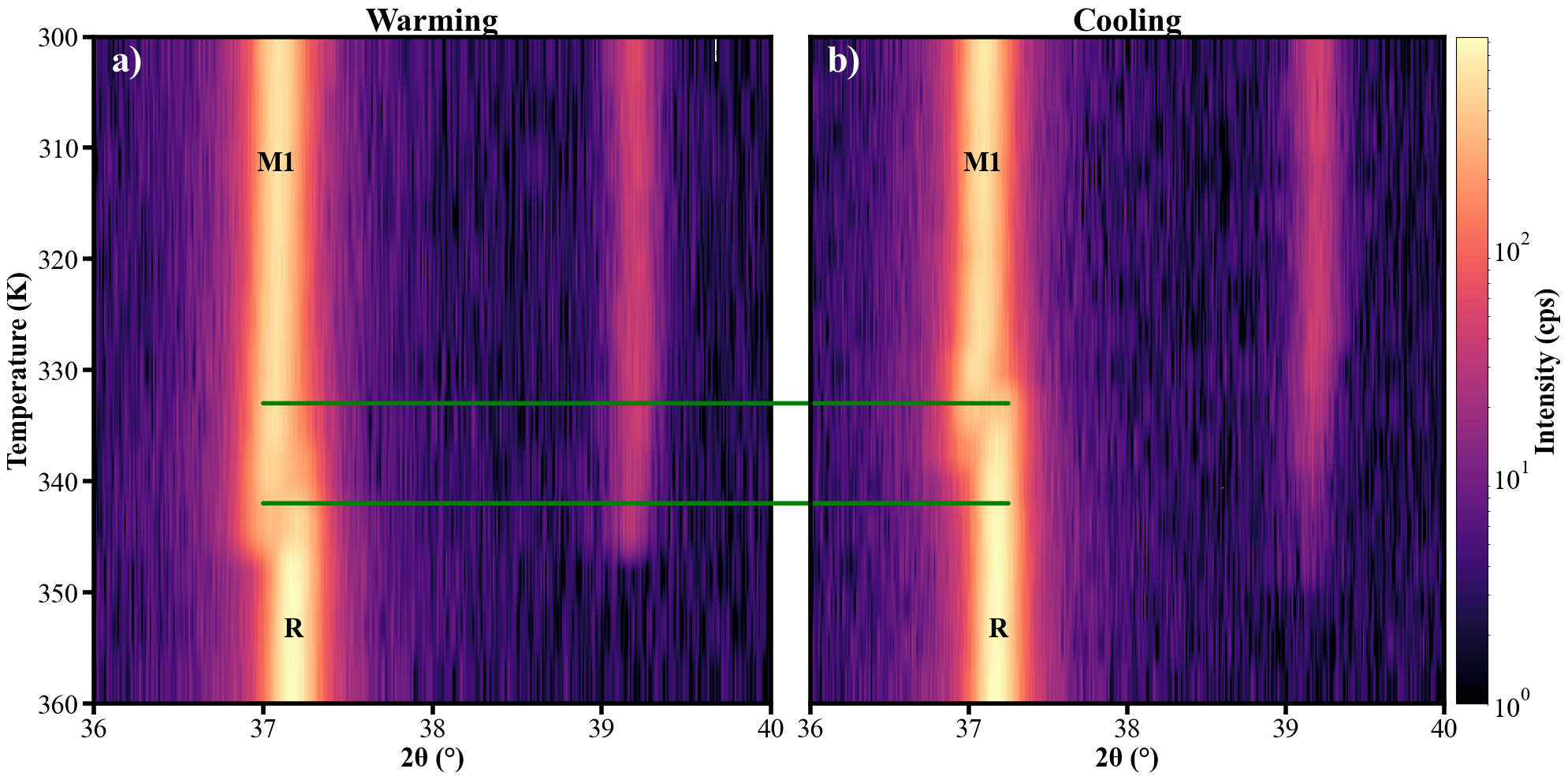}
\caption{\label{supfig:warmcool} a) Mapping of the warming cycle of the unpatterned film showing the temperature progression of the phase transformation.  
b) Mapping of the cooling cycle of the unpatterned film showing the temperature progression of the phase transformation.  
The green lines indicated the temperature width for this transition's hysteresis. 
The phases labeled correspond to the (200) peak for the M1 phase and the (011) for the R phase.
The peak to the right is the (210) monoclinic peak which travels out of alignment as temperature increases.}
\end{figure*}
\clearpage
\section{Heating Stage}

A heating stage using a Peltier heater was used to heat the sample close to the transition temperature. 
Previous measurements on the film in air showed no electrical differences when the sample was left above the transition temperature. 
Characterization of the heater showed that it was stable to $\pm$0.5 K over the course of 1 hour. 

\section{Preliminary Characterization}

\begin{figure*}
\includegraphics[width=\textwidth]{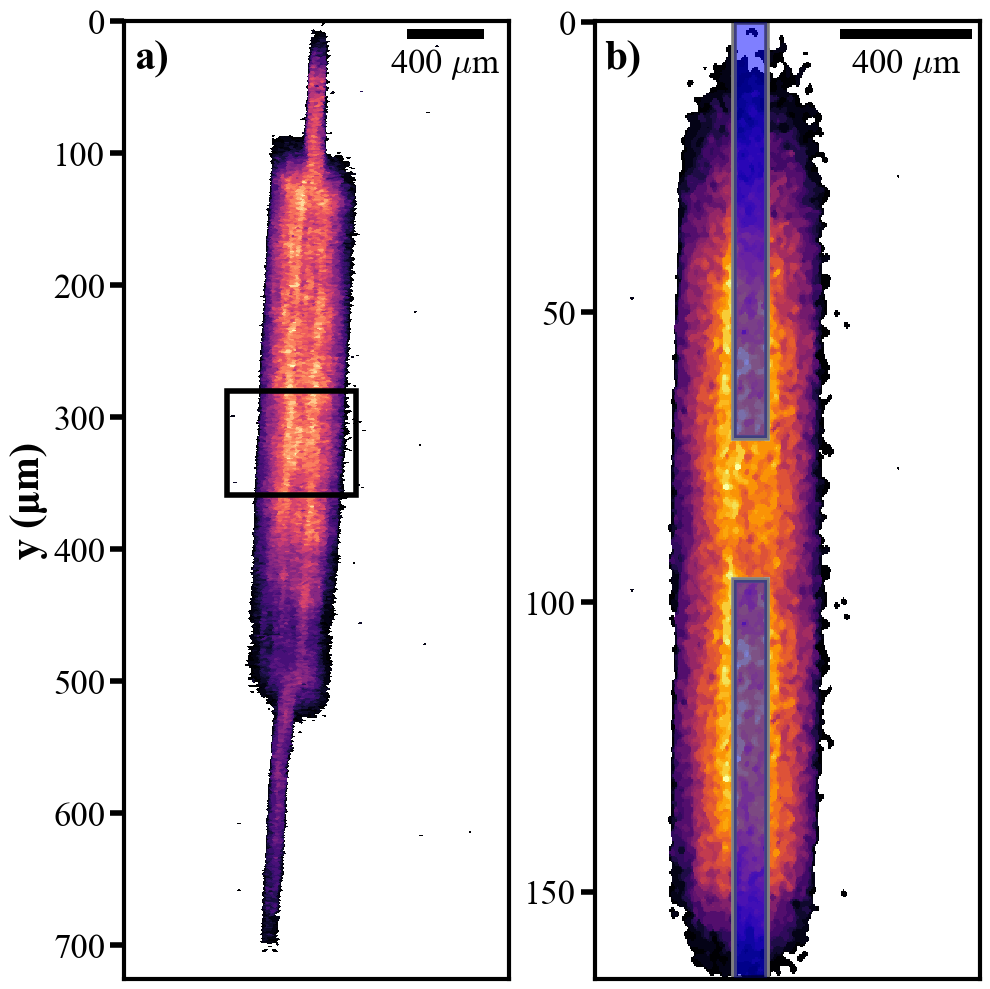}
\caption{\label{supfig:directDevice} a) Image of the full \VO device showing the 400x400 \textmu m\textsuperscript{2} pads together with the \VO underneath the electrodes leaving the pads. 
Scales for the horizontal direction, which is in the scattering plane are indicated by the top right scale bar while the vertical scales are given by the y-axis. 
The gold electrodes can be seen stretching from the top to the bottom with a gap in the center of the black square. 
b) A zoomed-in region of the black square shown in a) with electrodes indicated by the blue overlays. 
Scales are given in the same manner as a).}
\end{figure*}

Preliminary synchrotron measurements were conducted at 6-ID-C of the Advanced Photon Source (APS) which gave information on peak formation and preliminary imaging of the \VO device. 
The first images of the device (Fig. \ref{supfig:directDevice}) utilized a novel direct detection system with 8 \textmu m pixel sizes [45].
This allowed us to image the whole area of the \VO pad with device and electrodes around it, though the coarser resolution resulted from this setup makes it hard to study the spatial distribution of the M1 and R phases

\clearpage

\section{Image Processing Procedure}

Detector images were collected in 16-bit tif file and collected for every imaging scan performed. 
A background subtraction using a background estimator of a biweight background estimator as a robust measure of the background. 
This method gave a consistent background with dark images collected during one of the scans. 
To remove hot pixels, we utilized a median filter with a kernel size of 5. 
A bottom threshold application was also applied to remove any additional backgrounds missed near the edge of the detector. 
Images were converted to edf files for use with the \textit{darfix} package [46].

\clearpage
\section{Memory Reset and Device Cycling}

\begin{figure*}[h]
\includegraphics[width=\textwidth]{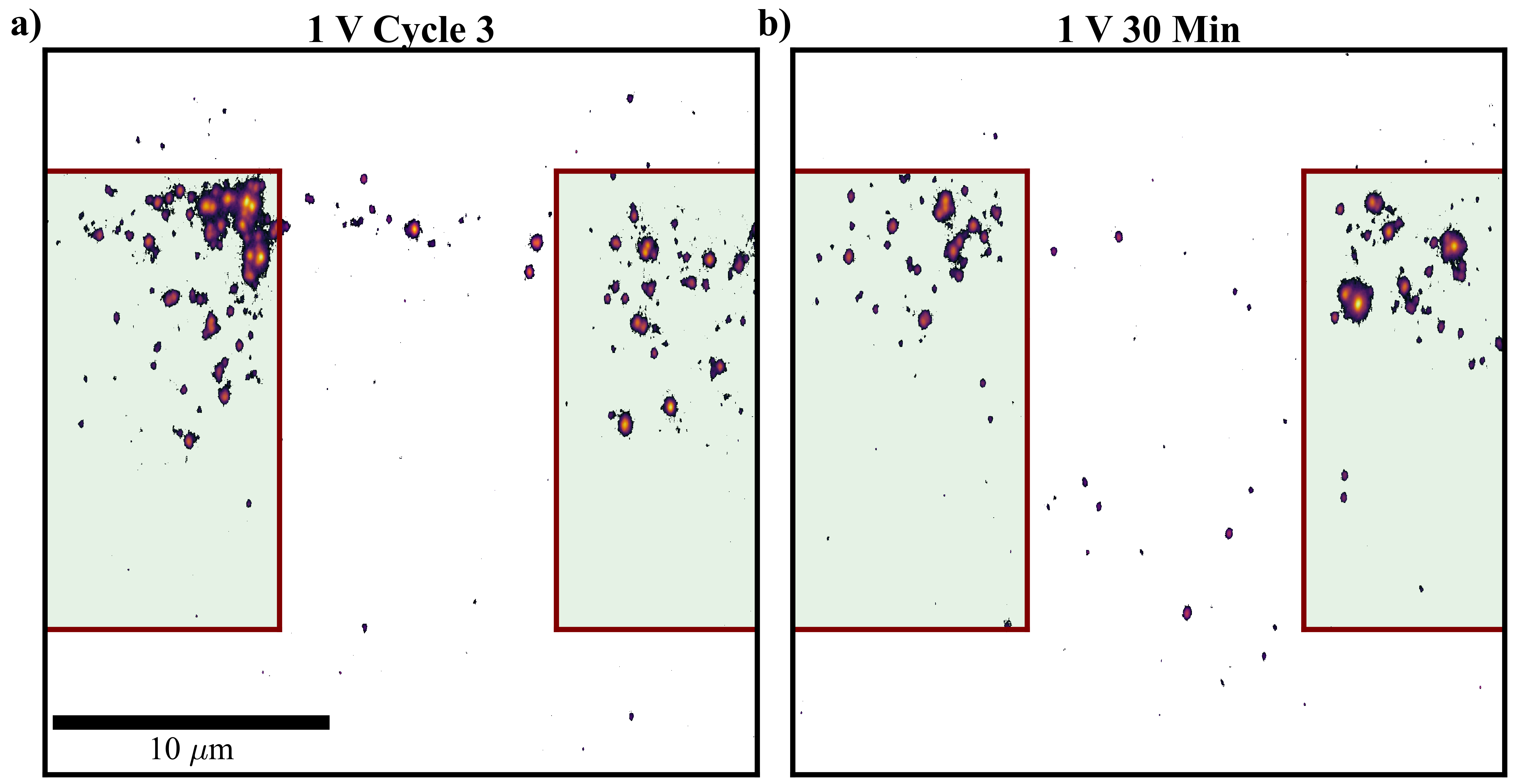}
\caption{\label{supfig:relax} a) Image taken at 1V at the beginning of the 3rd cycle on device 2. 
We continue to observe the R phase persistence near the top regions of the device prior to switching. 
b) An image showing the reset device after 30 minutes of relaxing showing much less phase persistence. 
There also appears a redistribution of the R phase domains beneath the electrode and an equal amount of R domains appearing at the top and bottom of the device in the gap.}
\end{figure*}

\begin{figure*}
\includegraphics[width=\textwidth]{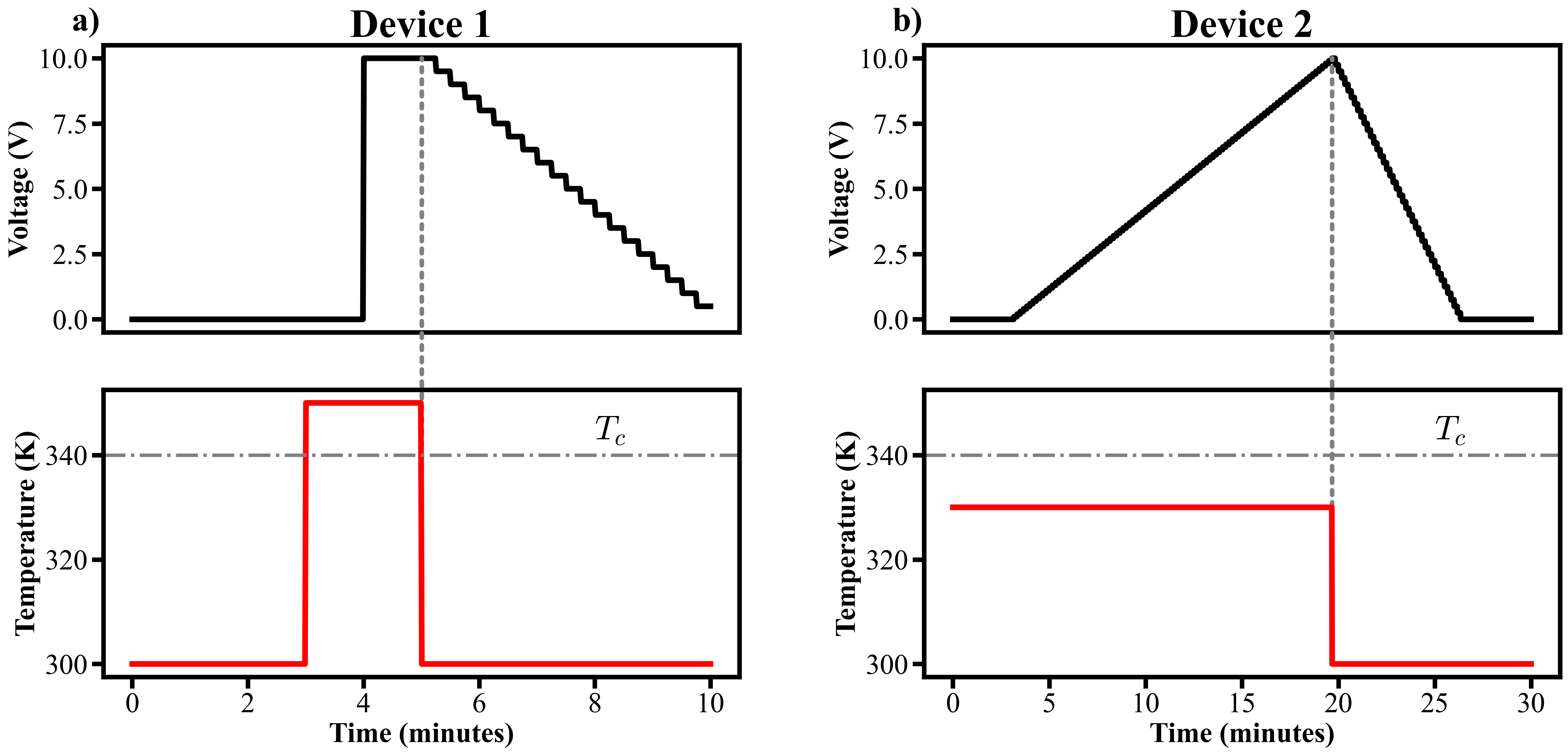}
\caption{\label{supfig:cyclingPlots} a) The associated temperature (red) and voltage (black) applications during the operation of device 1 where the stable filament forms. 
b) The associated temperature (red) and voltage (black) applications during the operation of device 2 showing the cycling process.}
\end{figure*}
\clearpage
\section{Device 1 Rastering}

\begin{figure*}[h]
\includegraphics[width=\textwidth]{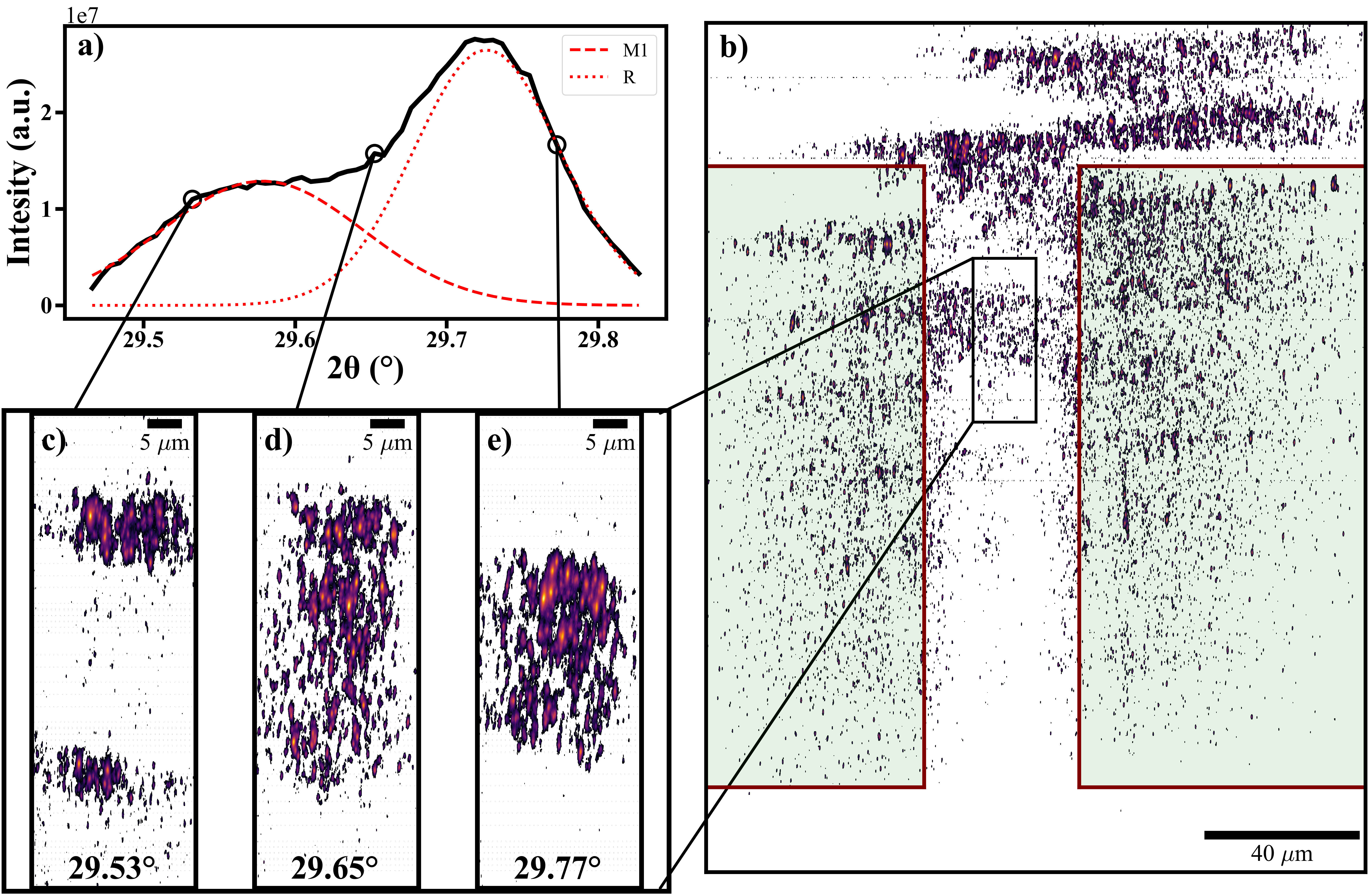}
\caption{\label{supfig:rastering} a) The integrated diffraction intensity of the images of the steady-state filament at 10 V. 
The two fitted gaussians correspond to the M1 and R phases associated with the insulating and metallic phases, respectively. 
b) A full device raster taken at the same diffraction condition as (e) of device 1. 
The small speckle regions are domains of the \VO film in the R phase. 
The red lines and green-shaded regions indicate the electrodes which were confirmed by an optical microscope on the setup. 
The black central box indicates where (c-e) were collected. 
There is significant R phase formation beneath the electrodes as well as outside the bounds of the device. 
c-e) DFXM images corresponding to the indicated points on (a). 
c) The M1 phase shows there is a boundary between the central filament and other filaments that form. 
d) An image of the same region where the diffraction conditions between the M1 and R phase are indistinguishable. 
e) An image of the R phase which corresponds to the deficit region of (c).}
\end{figure*}


\clearpage
\bibliography{VO2_Main}